\documentclass[pra,twocolumn,showpacs,superscriptaddress,10pt]{revtex4-1}
\usepackage[colorlinks,citecolor=blue,linkcolor=red,dvipdfm]{hyperref}
\usepackage{amsmath,amssymb,graphicx}

\begin{document}

\title{Families of fundamental and multipole solitons in a cubic-quintic
nonlinear lattice in fractional dimension}

\author{Liangwei Zeng}
\affiliation{College of Physics and Optoelectronic Engineering, Shenzhen University, Shenzhen 518060, China}
\affiliation{Shenzhen Key Laboratory of Micro-Nano Photonic Information Technology, College of Physics and
Optoelectronics Engineering, Shenzhen University, Shenzhen 518060, China}

\author{Dumitru Mihalache}
\affiliation{Horia Hulubei National Institute of Physics and Nuclear Engineering, Magurele, Bucharest, RO-077125, Romania}

\author{Boris A. Malomed}
\affiliation{Department of Physical Electronics, School of Electrical Engineering, Faculty of Engineering, and the Center for
Light-Matter Interaction, Tel Aviv University, P.O.B. 39040, Ramat Aviv, Tel Aviv, Israel}
\affiliation{Instituto de Alta Investigaci\'{o}n, Universidad de Tarapac\'{a}, Casilla 7D, Arica, Chile}

\author{Xiaowei Lu}
\affiliation{College of Physics and Optoelectronic Engineering, Shenzhen University, Shenzhen 518060, China}
\affiliation{Shenzhen Key Laboratory of Micro-Nano Photonic Information Technology, College of Physics and
Optoelectronics Engineering, Shenzhen University, Shenzhen 518060, China}

\author{Yi Cai}
\affiliation{College of Physics and Optoelectronic Engineering, Shenzhen University, Shenzhen 518060, China}
\affiliation{Shenzhen Key Laboratory of Micro-Nano Photonic Information Technology, College of Physics and
Optoelectronics Engineering, Shenzhen University, Shenzhen 518060, China}

\author{Qifan Zhu}
\affiliation{College of Physics and Optoelectronic Engineering, Shenzhen University, Shenzhen 518060, China}
\affiliation{Shenzhen Key Laboratory of Micro-Nano Photonic Information Technology, College of Physics and
Optoelectronics Engineering, Shenzhen University, Shenzhen 518060, China}

\author{Jingzhen Li}
\email{\underline{lijz@szu.edu.cn}}
\affiliation{College of Physics and Optoelectronic Engineering, Shenzhen University, Shenzhen 518060, China}
\affiliation{Shenzhen Key Laboratory of Micro-Nano Photonic Information Technology, College of Physics and
Optoelectronics Engineering, Shenzhen University, Shenzhen 518060, China}

\begin{abstract}
We construct families of fundamental, dipole, and tripole solitons in the
fractional Schr\"{o}dinger equation (FSE)\ incorporating self-focusing cubic
and defocusing quintic terms modulated by factors $\cos ^{2}x$ and $\sin
^{2}x$, respectively. While the fundamental solitons are similar to those in
the model with the uniform nonlinearity, the multipole complexes exist only
in the presence of the nonlinear lattice. The shapes and stability of all
the solitons strongly depend on the L\'{e}vy index (LI)\ that determines
the FSE fractionality. Stability areas are identified in the plane of LI and
propagation constant by means of numerical methods, and some results are
explained with the help of an analytical approximation. The stability areas
are broadest for the fundamental solitons and narrowest for the tripoles.
\\
\\
\textbf{key words:} Multipole solitons; Cubic-quintic nonlinear lattice; Fractional Schr\"{o}dinger equation
\end{abstract}

\maketitle

\section{Introduction}

It is commonly known that stable bright and dark solitons exist in
one-dimensional uniform self-focusing and self-defocusing nonlinear media,
respectively \cite{KivsharAgrawal,DS}. However, the situation may be totally
different in higher dimensions, where bright solitons are destabilized by
the collapse \cite%
{LB1998,STA,Malomed2005,Chen2012,review2016,review2019_1,review2019_2,SREV}.
One of solutions to this problem is offered by the use of spatially
inhomogeneous settings \cite{LL1,LL2,LL3,LL4,LL5,WGA}. First, linear
periodic potentials \cite{Brazhnyi,Morsch,Vysloukh,Vysloukh2,Yang,LPP},
which are induced by optical lattices in Bose-Einstein condensates (BECs),
or photonic lattices in optical waveguides, help to stabilize various types
of self-trapped modes, including fundamental \cite{LL6}, dipole \cite{LL7},
and vortex \cite{Baizakov,LL2,Baizakov2,LL8} solitons. Another option is
the use of periodic spatial modulation of the local nonlinearity strength,
alias nonlinear lattices \cite{NLREV}. Many types of solitons are maintained
in nonlinear lattices \cite{NL1,NL2,NL3}, \textit{viz}., fundamental ones,
as well as dipole, tripole, and vortex solitons. In addition, combinations
of linear and nonlinear lattices can be used to stabilize various species of
solitons \cite{CLN1,Sakaguchi,CLN2}. Inhomogeneous self-defocusing
nonlinearity \cite{SDN1,SDN2,Dror} is another setting in which many species
of stable self-trapped modes have been predicted, such as multipole \cite%
{SDN1,SDN2} and vortex \cite{SDN3,Tian,Wu,SDN4} solitons, vortex clusters
\cite{SDN5,SDN6}, skyrmions \cite{SDN7}, soliton gyroscopes \cite{SDN8}, and
flat-top solitons \cite{SDN9,SDN10}. Inhomogeneous nonlinearities can be
implemented by means of creation of a spatially nonuniform distribution of
dopants in photorefractive optical media \cite{FRM}. In BEC, similar setups
can be created too, using the Feshbach resonance controlled by spatially
nonuniform external fields \cite{FR1,FR2,FR3}.

In this work, we aim to address the possibility of using nonlinear lattices
as an ingredient of the fractional Schr\"{o}dinger equation (FSE), which was
introduced by Laskin two decades ago in the context of quantum mechanics
\cite{Lask1,Lask2,Lask3,EXP3}. The main parameter that defines the
fractionality of the FSE is the L\'{e}vy index (LI), $\alpha $, see Eq. (\ref%
{NLFSE}) below. The experimental implementation of the FSE in
condensed-matter \cite{EXP1,EXP2} and optical \cite{EXP4} setups, where
nonlinearity is a natural feature, has drawn interest to the possibility of
existence of solitons in fractional dimensions \cite{Frac1,LiRRP,Frac2,Frac3}%
. In particular, \textquotedblleft accessible solitons" \cite{Frac4,Frac5}
and self-trapped states of vectorial \cite{Frac6}, gap \cite{Frac7},
nonlocal \cite{Frac8}, vortical \cite{Frac9}, and multi-peak types \cite%
{Frac10} have been predicted in FSE models, as well as soliton clusters \cite%
{Frac11,Frac11b}, symmetry breaking of solitons \cite{Frac12,Frac13}, coupled
solitons \cite{Frac13b} and dissipative solitons in a fractional
complex-Ginzburg-Landau model \cite{Frac14}. In the case of the ubiquitous
cubic (Kerr) self-focusing, the solitons are unstable at $\alpha \leq 1$,
as the combination of such values with the Kerr nonlinearity gives rise to
the collapse.

The objective of this work is to identify domains of existence and stability
of fundamental and multipole (dipole and tripole) solitons in FSE with a
nonlinear lattice combining the cubic self-focusing and quintic defocusing
(vortices produced by the FSE including the uniform cubic-quintic (CQ)
nonlinearity were recently addressed in Ref. \cite{Frac9}). While
fundamental solitons produced by this model are not strongly different from
their counterparts in the one with the uniform nonlinearity, it opens a way
to create stable multipoles, i.e., complexes of solitons with opposite signs
placed at adjacent sites of the lattice.

The rest of the paper is organized as follows. In Sec. \ref{sec2} we present
the model and the method used for the linear-stability analysis. Then, in
Sec. \ref{sec3}, we summarize extensive numerical results that make it
possible to outline the stability domains for the fundamental solitons and
multipoles, \textit{viz}., dipole and tripole bound states. These results,
based on the computation of eigenvalues for small perturbations, are
corroborated by direct simulations of perturbed propagation dynamics of the
solitons. The paper is concluded in Sec. \ref{sec4}.

\section{The model}

\label{sec2} The FSE modeling the propagation of a light beam with field
amplitude $E$ under the action of the fractional diffraction in the medium
with the CQ\ spatially modulated nonlinearity is written in the normalized
form:
\begin{equation}
i\frac{\partial E}{\partial z}=\frac{1}{2}\left( -\frac{\partial ^{2}}{%
\partial x^{2}}\right) ^{\alpha /2}E-\mathrm{g}(x)\left\vert E\right\vert
^{2}E+\xi (x)\left\vert E\right\vert ^{4}E,  \label{NLFSE}
\end{equation}%
where $z$ is the propagation distance, while the fractional diffraction
operator with LI $\alpha $ is defined by the integral expression \cite%
{Lask1,Zhong},%
\begin{equation}
\begin{split}
\left( -\frac{\partial ^{2}}{\partial x^{2}}\right) ^{\alpha /2}E=& \frac{1}{%
2\pi }\int_{-\infty }^{+\infty }|p|^{\alpha }dp \\
\times & \int_{-\infty }^{+\infty }d\xi \exp \left[ ip\left( x-\xi \right) %
\right] E(\xi ),
\end{split}%
\end{equation}%
the limit of $\alpha =2$ corresponding to the usual second derivative. In
the case of the cubic-only self-focusing nonlinearity, LI is limited by
values $\alpha >1$, as the collapse occurs at $\alpha \leq 1$, as mentioned
above. However, the inclusion of the quintic term with the defocusing sign
suppresses the collapse and makes it possible to consider values $\alpha <1$
as well \cite{Frac9}, although the solutions tend to be unstable in the
latter case, see Fig. \ref{fig2}(b) below. Note that, in the case of
constant nonlinearity coefficients $\mathrm{g}$ and $\xi $, Eq. (\ref{NLFSE}%
) admits continuous-wave (CW) solutions with an arbitrary real wavenumber $k$
and arbitrary amplitude $E_{0}$:%
\begin{eqnarray}
E_{\mathrm{CW}} &=&E_{0}\exp \left( ib_{\mathrm{CW}}z+ikx\right) ,  \notag \\
b_{\mathrm{CW}} &=&-\frac{1}{2}|k|^{\alpha }+\mathrm{g}E_{0}^{2}-\xi
E_{0}^{4}.  \label{CW}
\end{eqnarray}

The nonlinear lattice is introduced, replacing the constant coefficients by
the spatially-modulated ones:
\begin{equation}
\mathrm{g}(x)=\mathrm{cos}^{2}x,\,\xi (x)=\mathrm{sin}^{2}x.  \label{CQNL}
\end{equation}%
Here, the lattice period and amplitudes are fixed, severally, to be $\pi $
and $1$, by means of rescaling. Stationary solutions with real propagation
constant $b$ are sought for as $E(x,z)=U(x)\mathrm{exp}(ibz)$ (cf. the CW
solution (\ref{CW})), where the stationary field, $U(x)$, satisfies the
following equation:
\begin{equation}
-bU=\frac{1}{2}\left( -\frac{\partial ^{2}}{\partial x^{2}}\right) ^{\alpha
/2}U-\left( \mathrm{cos}^{2}x\right) \left\vert U\right\vert ^{2}U+\left(
\mathrm{sin}^{2}x\right) \left\vert U\right\vert ^{4}U.  \label{NLFSES}
\end{equation}%
Soliton solutions to Eq. (\pageref*{NLFSES}) are characterized by their
integral power,
\begin{equation}
P=\int_{-\infty }^{+\infty }\left\vert U(x)\right\vert ^{2}dx.  \label{POWER}
\end{equation}%
Equation (\ref{NLFSES}) is solved below by dint of the modified
squared-operator method \cite{Lakoba}.

The linear-stability analysis will be performed for a perturbed solution
taken as
\begin{equation}
E=[U(x)+p(x)\mathrm{exp}(\lambda z)+q^{\ast }(x)\mathrm{exp}(\lambda ^{\ast
}z)]\mathrm{exp}(ibz),  \label{PERB}
\end{equation}%
where $p(x)$ and $q^{\ast }(x)$ represent the eigenmode of small
perturbations corresponding to eigenvalue $\lambda $ (the instability growth
rate). The eigenmodes are solutions of the linearized equations produced by
the substitution of ansatz (\ref{PERB}) in Eq. (\ref{NLFSE}):
\begin{gather}
i\lambda p=+\frac{1}{2}\left( -\frac{\partial ^{2}}{\partial x^{2}}\right)
^{\alpha /2}p+bp-\left( \mathrm{cos}^{2}x\right) U^{2}(2p+q)  \notag
\label{LAS} \\
+\left( \mathrm{sin}^{2}x\right) U^{4}(3p+2q),  \notag \\
\\
i\lambda q=-\frac{1}{2}\left( -\frac{\partial ^{2}}{\partial x^{2}}\right)
^{\alpha /2}q-bq+\left( \mathrm{cos}^{2}x\right) U^{2}(2q+p)  \notag \\
-\left( \mathrm{sin}^{2}x\right) U^{4}(3q+2p).  \notag
\end{gather}%
The solitons are stable if $\mathrm{Re}(\lambda )=0$ for all eigenvalues.
The so predicted (in)stability will be verified through direct numerical
simulations of the full evolution equation (\ref{NLFSE}), performed by means
of the split-step Fourier method.

\begin{figure}[tbp]
\begin{center}
\includegraphics[width=1\columnwidth]{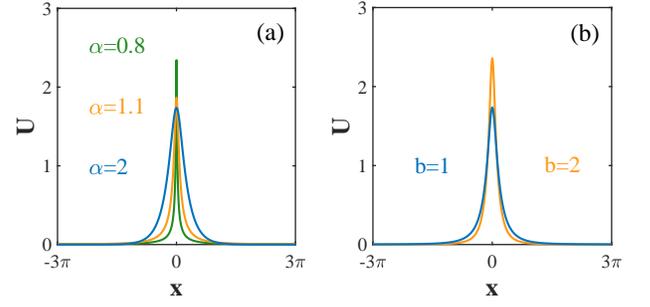}
\end{center}
\caption{Profiles of fundamental solitons: (a) with different values of LI (L%
\'{e}vy index) $\protect\alpha $, for propagation constant $b=1$; (b) with
different values of $b$, for $\protect\alpha =1.5$. Note that $\protect%
\alpha =2$ corresponds to the usual local (non-fractional) Schr\"{o}dinger
equation.}
\label{fig1}
\end{figure}

\begin{figure}[tbp]
\begin{center}
\includegraphics[width=1\columnwidth]{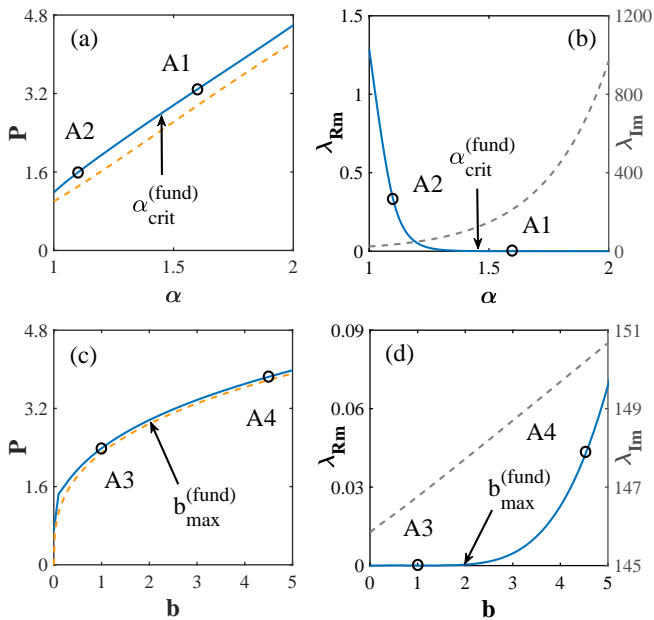}
\end{center}
\caption{Soliton's power $P$ (a) and the maximum real part of eigenvalues
(instability growth rate) $\protect\lambda _{\mathrm{Rm}}$ (b) versus $%
\protect\alpha $ for fundamental solitons at $b=2$. In panel (b), the
vertical arrow indicates the boundary of the stabillity region, with very
weak instability to the left of it. Power $P$ (c) and $\protect\lambda _{%
\mathrm{Rm}}$ (d) versus $b$ for fundamental solitons at $\protect\alpha =1.5
$. The arrow in panel (d) indicates the respective boundary of the stability
region. Panels (a) and (c) include the best-fit approximation (red dashed
lines) provided by Eq. (\protect\ref{scaling}), with $\mathrm{const}=9$ and $%
12$, respectively. Panels (b) and (d) include the imaginary part, $\protect%
\lambda _{\mathrm{Im}}$ (gray dashed lines), corresponding to $\protect%
\lambda _{\mathrm{Rm}}$. Simulations of perturbed evolution of the
fundamental solitons marked by labels A1--A4 are displayed in Figs. \protect
\ref{fig3}(a-d), respectively.}
\label{fig2}
\end{figure}

\begin{figure}[tbp]
\begin{center}
\includegraphics[width=1\columnwidth]{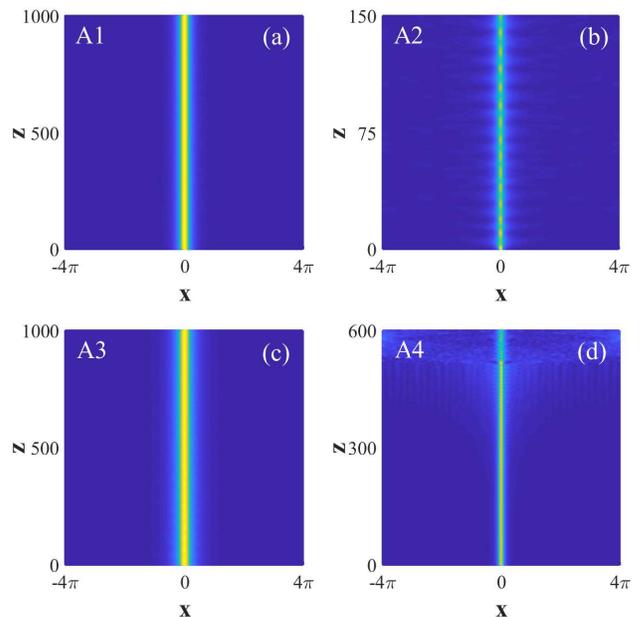}
\end{center}
\caption{The propagation of perturbed fundamental solitons: (a) a stable
soliton with $\protect\alpha =1.6$, $b=2$; (b) an unstable onen with $%
\protect\alpha =1.1$, $b=2$; (c) a stable soliton with $\protect\alpha =1.5$%
, $b=1$; (d) an unstable one with $\protect\alpha =1.5$, $b=4.5$.}
\label{fig3}
\end{figure}

\section{Numerical results}

\label{sec3} In this Section, we report numerical findings for solitons in
the FSE with the CQ lattice, including their profiles, stability domains,
and perturbed propagation. The Section is divided in two parts, which
address, separately, fundamental and multipole (dipole and tripole) solitons.

\subsection{Fundamental solitons}

Typical profiles of fundamental stationary solitons are presented in Fig. %
\ref{fig1}. Naturally, the maximum of the local intensity is located at $x=0$%
, where the self-focusing and defocusing coefficients in Eq. (\ref{NLFSES})
attain, respectively, a maximum and minimum (zero). Accordingly, the soliton
is stabilized by being placed at the bottom of an effective nonlinear
potential well induced by the spatially-modulated CQ nonlinearity. In fact,
Fig. \ref{fig1} demonstrates that the fundamental solitons are completely
confined to a spatial interval $|x|<3\pi /4$, hence Eq. (\ref{NLFSES})
suggests that the shape of the solitons is chiefly determined by the cubic
term. Then, taking, for an approximate estimate, a constant value of the
respective coefficient, $\mathrm{g}$ (see Eq. (\ref{NLFSE})), Eq. (\ref%
{NLFSES}) predicts a scaling relation between the propagation constant,
amplitude, and width ($W$): $b\sim W^{-\alpha }\sim A^{2}$, which determines
the approximate scaling of the integral power,
\begin{equation}
P\sim A^{2}W\sim \left( \mathrm{const}\cdot b\right) ^{1-1/\alpha }.
\label{scaling}
\end{equation}

As LI $\alpha $ increases, the amplitude $A$ of the fundamental soliton
decreases, while its width increases, as seen in Fig. \ref{fig1}(a). In this
figure, the amplitudes are $A(\alpha =0.8)=2.34$, $A(\alpha =1.1)=1.86$, and
$A(\alpha =2)=1.73$. On the other hand, it is natural that the amplitude
increases and width decreases with the growth of propagation constant $b$,
as shown in Fig. \ref{fig1}(b), in agreement with the above-mentioned
scaling relations. In the latter figure, the amplitudes of the fundamental
soliton are $A(b=1)=1.73$ and $A(b=2)=2.36$.

Next, we present the characteristics of families of fundamental solitons. First,
in Fig. \ref{fig2}(a) we display the relation between the soliton's power $P$%
, calculated as per Eq. (\ref{POWER}), and LI $\alpha $, for a fixed value
of the propagation constant. It is seen that the dependence $P(\alpha )$ is
a nearly linear one. In fact, this dependence may be approximately explained
by Eq. (\ref{scaling}), as shown by the fitting line in Fig. \ref{fig2}(a).

The stability is determined by the dependence of the maximum real part of
eigenvalues, $\lambda _{\mathrm{Rm}}$, obtained from the numerical solution
of Eq. (\ref{LAS}), on $\alpha $, as shown in Fig. \ref{fig2}(b), which
identifies the respective stability region
\begin{equation}
\alpha > \alpha _{\mathrm{crit}}^{\mathrm{(fund)}}\approx 1.45,
\label{alpha-fund}
\end{equation}%
as defined by condition $\lambda _{\mathrm{Rm}}=0$. As mentioned above, the
plots in Figs. \ref{fig2}(a) and (b) can be extended towards $\alpha <1$,
but in that area the fundamental solitons are definitely unstable, as
suggested by Fig. \ref{fig2}(b).

It is well known that a necessary (but, generally, not sufficient) stability
condition for bright solitons is provided by the Vakhitov-Kolokolov (VK)
criterion, which is formulated in terms of the dependence of the integral
power on the propagation constant: $dP/db>0$ \cite{Vakh,LB1998}. The relation
$P(b)$ is displayed in Fig. \ref{fig2}(c), which clearly complies with the
VK criterion at all values of $b$. The $P(b)$ dependences can be also
explained in a quasi-analytical form, making use of scaling relation (\ref%
{scaling}), as shown by means of fitting line in Fig. \ref{fig2}(c). In
particular, for the value of $\alpha =1.5$ adopted in Fig. \ref{fig2}(c) Eq.
(\ref{scaling}) yields $P\sim b^{1/3}$.

\begin{figure}[!tbp]
\begin{center}
\includegraphics[width=1\columnwidth]{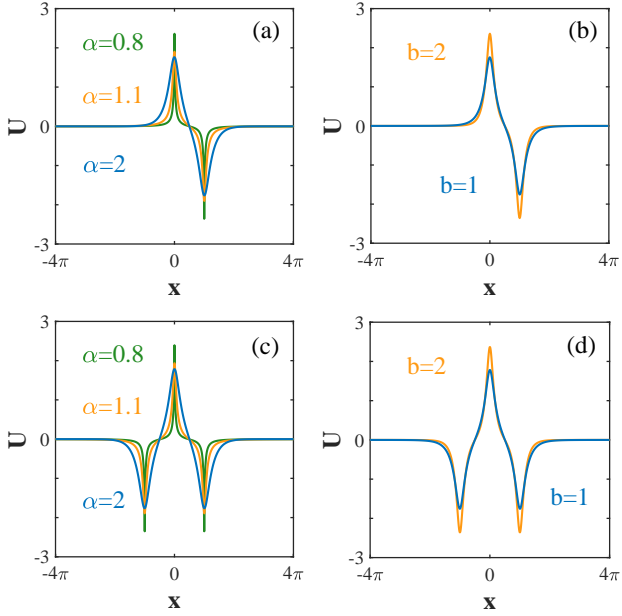}
\end{center}
\caption{Typical profiles of dipole bound states: (a) with different values
of $\protect\alpha $ at $b=1$; (b) with different values of $b$ at $\protect%
\alpha =1.8$. Profiles of tripoles: (c) with different values of $\protect%
\alpha $ at $b=1$; (d) with different values of $b$ at $\protect\alpha =1.8$%
. }
\label{fig4}
\end{figure}

\begin{figure}[!tbp]
\begin{center}
\includegraphics[width=1\columnwidth]{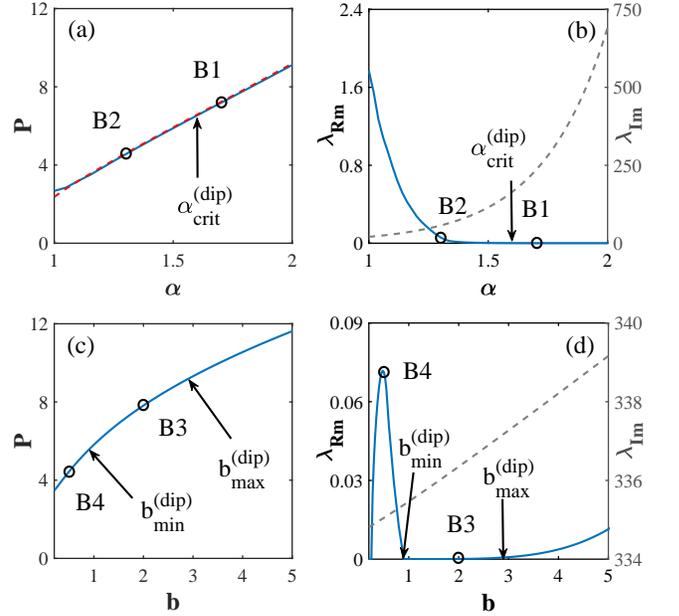}
\end{center}
\caption{Dipole bound states. Integral power $P$ (a) and the maximum real
part of perturbation eigenvalues $\protect\lambda _{\mathrm{Rm}}$ (b) versus
$\protect\alpha $ at a fixed propagation constant, $b=2$. Power $P$ (c) and $%
\protect\lambda _{\mathrm{Rm}}$ (d) versus $b$ for at fixed LI, $\protect%
\alpha =1.8$. Similar to Fig. \protect\ref{fig2}, arrows in panels (b) and
(d) indicate boundaries of the respective stability regions, the instability
being very weak to the left of the boundary in (b). A cardinal difference
from the fundamental solitons (cf. Fig. \protect\ref{fig2}(d)) is that the
stability region for the dipoles in panel (d) is bounded on the side of
small $b$. The red dashed line in (a) displays the data
for the fundamental solitons from Fig. \ref{fig2}(a), with the same
propagation constant, $b=2$, multiplied by $2$. Panels (b) and
(d) also include the imaginary part $\protect\lambda _{\mathrm{Im}}$
corresponding to $\protect\lambda _{\mathrm{Rm}}$ (gray dashed lines).
Perturbed propagation of the dipoles marked by B1--B4 are displayed in
Figs. \protect\ref{fig7}(a-d), respectively.}
\label{fig5}
\end{figure}

The stability analysis is continued by data displayed in Fig. \ref{fig2}(d),
which displays numerically exact results for the maximum instability growth
rate, $\lambda _{\mathrm{Rm}}(b)$, in the same interval of $b$ that is
shown in Fig. \ref{fig2}(c). It is seen that the stability actually takes
place in a part of the interval,
\begin{equation}
0\leq b<b_{\max }^{\mathrm{(fund)}}\approx 2,  \label{b-fund}
\end{equation}%
and unstable at $b>b_{\max }^{\mathrm{(fund)}}$. The instability of the
narrow solitons corresponding to large $b$ may be explained by the fact that
the effectively nonlocal fractional-diffraction operator cannot provide the
stability of very narrow modes, unlike the usual local operator
corresponding to $\alpha =2$.

Predictions for the stability produced by the computation of $\lambda _{%
\mathrm{Rm}}$ have been verified by direct simulations of the evolution of
the fundamental solitons performed in the framework of the full nonlinear
equation (\ref{NLFSE}), as shown in Fig. \ref{fig3}. In the simulations,
random initial perturbations at the $1\%$ amplitude level were added to the
input. In Fig. \ref{fig3} the left and right column display the stable and
unstable propagation, respectively. In the former case, the solitons with $%
\lambda _{\mathrm{Rm}}=0$ demonstrate completely stable evolution up to $%
z=1000$, which corresponds, with $A\simeq 2$ (see Fig. \ref{fig1}) to $\sim
5000$ characteristic nonlinearity lengths, $z_{\mathrm{nonlin}}\sim A^{-2}$.
On the other hand, unstable solitons with $\lambda _{\mathrm{Rm}}>0$
oscillate irregularly in Fig. \ref{fig3}(b) or very slowly decay through the
emission of small-amplitude \textquotedblleft radiation" in Fig. \ref{fig3}%
(d). In fact, even the unstable solitons may be categorized as objects that
will seem sufficiently robust ones in the experiment, as the instability
development does not destroy them. Additional simulations demonstrate that
the instability, if any, keeps the same mild character at other values of
the parameters.

To conclude this subsection, it is relevant to mention that the simplified
model, without the quintic term and a constant coefficient in front of the
cubic one, provide similar results for the shape and stability of the
fundamental solitons. Howvere, such a model is not able to produce dipole
and tripole modes presented below.

\subsection{Dipoles and tripoles}

As mentioned above, the nonlinear lattice offers the possibility to create
multipole bound states. Typical profiles of dipole and tripole complexes,
with opposite signs of adjacent fundamental solitons that build the
dipoles, are presented, severally, in Figs. \ref{fig4}(a,b) and (c,d). The
dipoles are composed, roughly speaking, of two fundamental solitons placed
at adjacent local minima of the potential energy, $x=0$ and $\pi $ (see Eq. (%
\ref{CQNL})). Dipoles of this type are usually called intersite-centered
(alias densely-packed) ones, without an empty site in the middle, which
would be seen in \textit{onsite-centered} modes. They may be stable due to
the balance between repulsion between the constituents with opposite signs
and pinning to the underlying lattice \cite{Kapitula}. On the other hand,
lattice-based complexes built of adjacent fundamental solitons with
identical signs are unstable in accordance with the general analysis \cite%
{Kapitula}, therefore they are not reported here. The tripoles are also
built as tightly-packed complexes.

According to Figs. \ref{fig4}(a) and (c), amplitudes and widths of both the
dipole and tripole families gradually decrease and increase, respectively,
with the growth of LI $\alpha $, similar to what is observed above for the
fundamental solitons, cf. Fig. \ref{fig1}(a). For example, the dipole's
amplitudes are $A(\alpha =0.8)=2.36$, $A(\alpha =1.1)=1.9$ and $A(\alpha
=2)=1.76$. Further, it is seen in Figs. \ref{fig4}(b) and (d) that the
amplitude and width, respectively, gradually increase and decrease with the
growth of propagation constant $b$, also similar to properties of the
fundamental solitons shown in Fig. \ref{fig1}(b). Accordingly, typical
values of the dipole's amplitude are $A(b=1)=1.75$, and $A(b=2)=2.36$.

\begin{figure}[!tbp]
\begin{center}
\includegraphics[width=1\columnwidth]{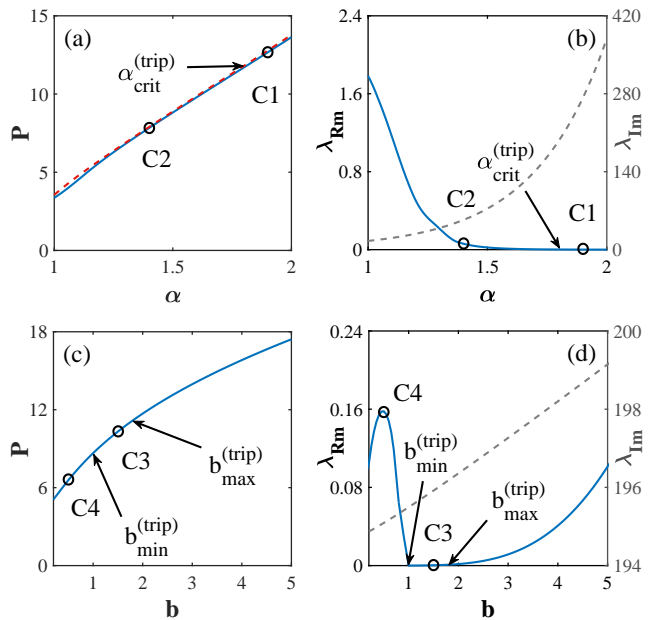}
\end{center}
\caption{Characteristics of the tripole complexes. Integral power $P$ (a)
and the maximum real part of perturbation eigenvalues $\protect\lambda _{%
\mathrm{Rm}}$ (b) versus $\protect\alpha $ at a fixed propagation constant, $%
b=2$. Power $P$ (c) and $\protect\lambda _{\mathrm{Rm}}$ (d) versus $b$ at
fixed LI, $\protect\alpha =1.8$. Similar to Fig. \protect\ref{fig5}, arrows
in (b) and (d) indicate boundaries of the respective stability regions.
The red dashed line in (a) displays the data
for the fundamental solitons from Fig. \ref{fig2}(a), with the same
propagation constant, $b=2$, multiplied by $3$.
Panels (b) and (d) also include the  imaginary
part $\protect\lambda _{\mathrm{Im}}$ (gray dashed lines) corresponding to $%
\protect\lambda _{\mathrm{Rm}}$. Simulations of the perturbed propagations
of the tripole states marked by C1--C4 are displayed in Figs. \protect\ref%
{fig8}(a-d), respectively.}
\label{fig6}
\end{figure}

\begin{figure}[!tbp]
\begin{center}
\includegraphics[width=1\columnwidth]{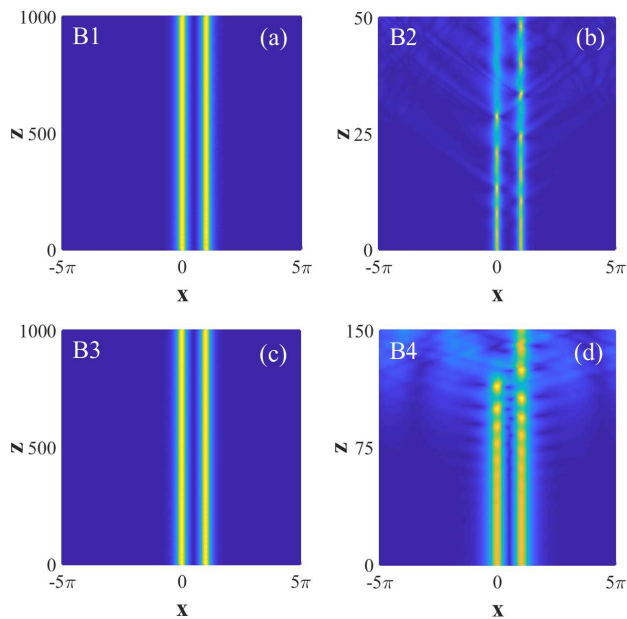}
\end{center}
\caption{The propagation of perturbed dipole solitons: (a) a stable soliton
at $\protect\alpha =1.7$, $b=2$; (b) an unstable one at $\protect\alpha =1.3$%
, $b=2$; (c) a stable soliton at $\protect\alpha =1.8$, $b=2$; (d) an
unstable one at $\protect\alpha =1.8$, $b=0.5$.}
\label{fig7}
\end{figure}

\begin{figure}[!h]
\begin{center}
\includegraphics[width=1\columnwidth]{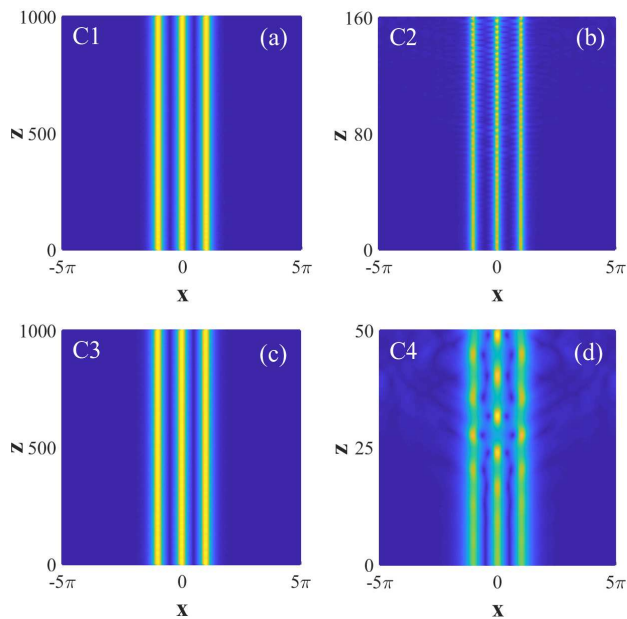}
\end{center}
\caption{The propagation of perturbed tripole solitons: (a) a stable one at $%
\protect\alpha =1.9$, $b=2$; (b) an unstable soliton at $\protect\alpha =1.4$%
, $b=2$; (c) a stable one at $\protect\alpha =1.8$, $b=1.5$; (d) an unstable
soliton at $\protect\alpha =1.8$, $b=0.5$.}
\label{fig8}
\end{figure}

Proceeding to characteristics of the family of dipole states, the respective
relation between the integral power $P$, as defined by Eq. (\ref{POWER}),
and LI $\alpha $ is displayed in Fig. \ref{fig5}(a), which shows a linear
growth of $P$ with $\alpha $. In fact, this $P(\alpha )$ curve is
nearly identical to its counterpart for the fundamental solitons, displayed
for the same fixed value of the propagation constant, $b=2$, in Fig. \ref{fig2}(a),
multiplied by $2$. Similar to the case of the fundamental
solitons (cf. Fig. \ref{fig2}(a)), this dependence can be explained,
although with poorer accuracy, by scaling relation (\ref{scaling}), as shown
by the fitting curve in \ref{fig5}(a). Next, the maximum real part of the
perturbation eigenvalues for the dipoles, $\lambda _{\mathrm{Rm}}$, produced
by the numerical solution of Eq. (\ref{LAS}), is displayed, as a function of
$\alpha $ in Fig. \ref{fig5}(b), demonstrating the stability region
\begin{equation}
\alpha >\alpha _{\mathrm{crit}}^{\mathrm{(dip)}}\approx 1.6,
\label{alpha-dip}
\end{equation}%
which is slightly narrower than its counterpart for the fundamental solitons
given by Eq. (\ref{alpha-fund}).

Another characteristic of the dipole family is $P(b)$ dependence, displayed
in versus $b$ in Fig. \ref{fig5}(c), which satisfies the VK criterion. Note
that the $P(b)$ dependence is still very accurately approximated by scaling
relation (\ref{scaling}), as shown by the fitting curve in the figure. The
corresponding dependence of the instability growth rate, $\lambda _{\mathrm{%
Rm}}$, on $b$ is displayed in Fig. \ref{fig5}(d). This plot demonstrates a
drastic difference from its counterpart for fundamental solitons in Fig. \ref%
{fig2}(d), as the stability interval for the dipoles is a \emph{window},
bounded both from above and from below,
\begin{equation}
b_{\min }^{\mathrm{(dip)}}\approx 0.9<b<b_{\max }^{\mathrm{(dip)}}\approx
2.9,  \label{b-dip}
\end{equation}%
cf. Eq. (\ref{b-fund}). The instability of broad dipoles corresponding to
small values of the propagation constant, $b<b_{\min }^{\mathrm{(dip)}}$, is
explained by the fact that the strong overlap between broad constituent
solitons that build the dipole leads to strong interaction between them,
which produces a destabilizing effect.

For the tripole complexes the relation between the integral power, $P$, and
LI $\alpha $ is presented in Fig. \ref{fig6}(a). This nearly linear
$P(\alpha )$ dependence is very close to its counterpart for the fundamental
solitons, displayed for the same fixed $b=2$, in Fig. \ref{fig2}(a),
multiplied by $3$ (again satisfying the VK criterion), similar to the above
notice that the $P(\alpha )$ curve in Fig. \ref{fig5}(a) is closely
approximated by its counterpart for the fundamental solitons from Fig. \ref%
{fig2}(a) multiplied by $3$. Next, in Fig. \ref%
{fig6}(b) the maximum real part of eigenvalues $\lambda _{\mathrm{Rm}}$,
produced by the numerical solution of Eq. (\ref{LAS}), is displayed as a
function of $\alpha $, the respective stability region being%
\begin{equation}
\alpha >\alpha _{\mathrm{crit}}^{\mathrm{(trip)}}\approx 1.8,
\label{alpha-trip}
\end{equation}%
which is narrower than its counterpart (\ref{alpha-dip}) for the dipoles.
The same eigenvalue is plotted versus the propagation constant
$b$ in Fig. \ref{fig6} (d). In terms of $b$, the stability window is
\begin{equation}
b_{\min }^{\mathrm{(trip)}}\approx 1<b<b_{\max }^{\mathrm{(trip)}}\approx
1.8,  \label{b-trip}
\end{equation}%
which is considerably smaller than a similar one (\ref{b-dip}) for the
dipoles. It is not surprising that the more complex bound states (tripoles)
is more fragile than simpler ones, arranged as dipoles.

Direct simulations of the perturbed propagation of dipole states are
presented in Fig. \ref{fig7}, in which the left and right columns refer,
respectively, to the stable and unstable propagation. According to the left
column, stable dipoles keep their shapes in the course of the very long
propagation, up to $z=1000$. On the other hand, those dipoles that are
predicted to be unstable by means of the linear-stability analysis feature
irregular oscillations and gradual decay through the emission of
small-amplitude waves, in the right column of Fig. \ref{fig7}. Similarly, in
Fig. \ref{fig8}, stable tripoles keep their shapes in the left column, while
their unstable counterparts also demonstrate irregular oscillations and
gradual delay.

\section{Conclusion}

\label{sec4} In this work, a one-dimensional model combining the fractional
diffraction and nonlinear CQ (cubic-quintic) lattice is introduced, and
families of fundamental and multipole (dipole and tripole) solitons in it
are reported. The shapes and stability of these states are established by
means of the systematic numerical investigation. Some results are explained
with the help of the analytical approximation, which is based on the
consideration of scaling properties of the underlying equation. The solitons
are stable above a critical value of the LI (L\'{e}vy index), $\alpha $. In
terms of the propagation constant, $b$, the fundamental solitons are stable
below a critical value, while the dipole and tripole complexes are stable in
intervals of $b$ limited from above and below. The stability areas of all
the states, predicted by the calculation of the perturbation eigenvalues,
are corroborated by direct simulations of the perturbed evolution. In the
simulations, the instability of some states seems weak, which suggests that
they may also represent relevant objects for experimental observation.
Naturally, the stability regions, in terms of both $\alpha $ and $b$, are
widest for the fundamental solitons, and narrowest for the most complex
states, \textit{viz., the tripoles.}

As an extension of the present analysis, it may be interesting to consider
mobility of the solitons on top of the underlying nonlinear lattice, and
collisions between moving ones. A straightforward extension may also address
bound states of more than three solitons, and dynamics of multi-soliton
arrays pinned to the nonlinear CQ lattice.

\section*{Declaration of Competing Interest}
The authors declare that they have no known competing financial interests or personal relationships that could have appeared to influence the work reported in this paper.

\section*{Funding}
National Major Instruments and Equipment Development Project of National Natural Science Foundation of China (No. 61827815); National Natural Science Foundation of China (No. 62075138); Science and Technology Project of Shenzhen (JCYJ20190808121817100, JCYJ20190808164007485); Israel Science Foundation (grant No. 1286/17).

\end{document}